\documentclass[acmtog]{acmart}
\acmSubmissionID{1240}

\usepackage{graphicx}
\usepackage{epstopdf}
\usepackage{enumitem}
\usepackage{color}
\usepackage{colortbl}
\usepackage{xcolor}
\usepackage{hyperref}
\usepackage{wrapfig}

\hypersetup{
	colorlinks=true,
	linkcolor=purple,
	filecolor=magenta,      
	urlcolor=cyan,
}

\usepackage[ruled]{algorithm2e} 

\SetAlFnt{\small}
\SetAlCapFnt{\small}
\SetAlCapNameFnt{\small}
\SetAlCapHSkip{0pt}

\setcopyright{rightsretained}
\acmJournal{TOG}
\acmYear{2024}
\acmVolume{43}
\acmNumber{4}
\acmArticle{90}
\acmMonth{7}
\acmDOI{10.1145/3658235}

\let\oldtwocolumn\twocolumn
\renewcommand\twocolumn[1][]{%
    \oldtwocolumn[{#1}{
    \begin{center}
           \includegraphics[width=\textwidth]{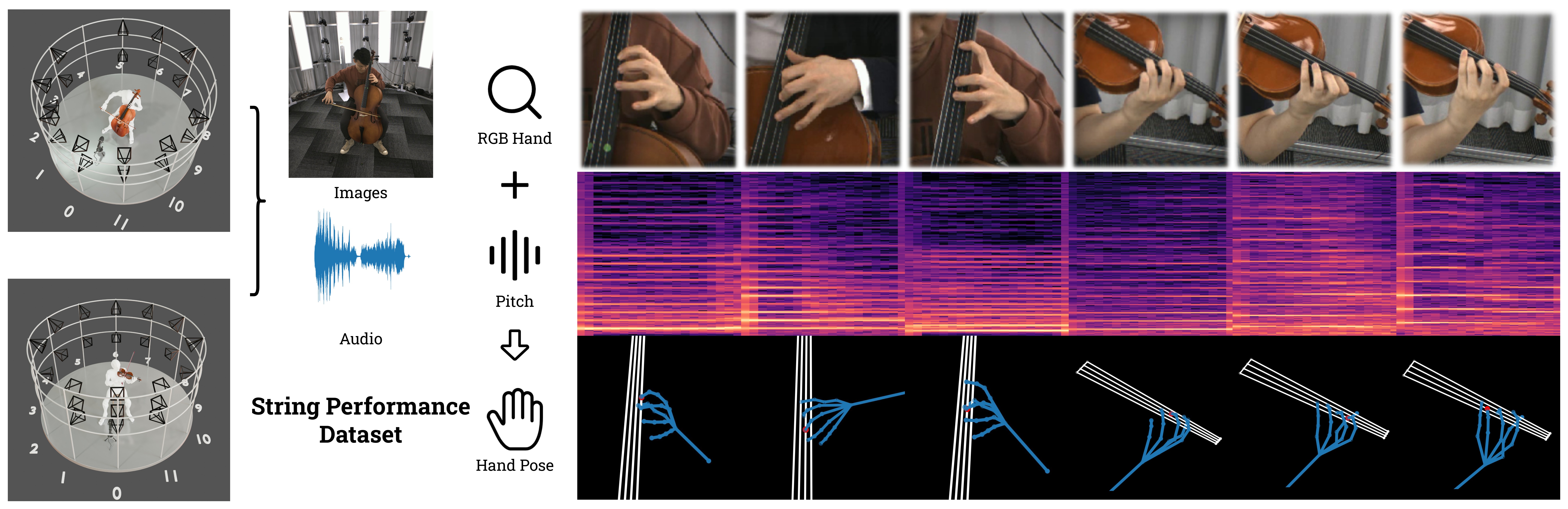}
           \captionof{figure}{We present the String Performance Dataset with an audio-guided multi-modal framework enhancing markerless motion capture for string performance.}
           \label{fig:1}
        \end{center}
    }]
} 
\begin{document}

\title{Audio Matters Too! Enhancing Markerless Motion Capture with Audio Signals for String Performance Capture}

\author{Yitong Jin}
\authornote{Equal Contribution}
\orcid{0009-0002-3979-5878}
\affiliation{%
 \institution{Central Conservatory of Music}
 \country{China}}
\affiliation{%
 \institution{Tsinghua University}
 \country{China}}
\email{jinyitong@mail.ccom.edu.cn}

\author{Zhiping Qiu}
\authornotemark[1]
\orcid{0009-0006-8663-1955}
\affiliation{%
 \institution{Central Conservatory of Music}
 \country{China}}
\affiliation{%
 \institution{Tsinghua University}
 \country{China}}
\email{zhiping_qiu@mail.ccom.edu.cn}

\author{Yi Shi}
\orcid{0000-0001-7500-192X}
\affiliation{%
 \institution{Central Conservatory of Music}
 \country{China}}
\affiliation{%
 \institution{Tsinghua University}
 \country{China}}
\email{shiyi@mail.ccom.edu.cn}

\author{Shuangpeng Sun}
\orcid{0009-0003-7988-7328}
\affiliation{%
 \institution{Tsinghua University}
 \country{China}}
\email{pengcheng786@gmail.com}

\author{Chongwu Wang}
\orcid{0009-0002-1384-941X}
\affiliation{%
 \institution{Central Conservatory of Music}
 \country{China}}
\email{1225@ccom.edu.cn}

\author{Donghao Pan}
\orcid{0009-0008-3863-286X}
\affiliation{%
 \institution{Central Conservatory of Music}
 \country{China}}
\email{pdh0227@sina.com}

\author{Jiachen Zhao}
\orcid{0000-0002-2864-9229}
\affiliation{%
 \institution{Tsinghua University}
 \country{China}}
\email{zhao_jiachen@163.com}

\author{Zhenghao Liang}
\orcid{0009-0008-7467-4469}
\affiliation{%
 \institution{Weilan Tech}
 \country{China}}
\email{liangzhenghaothu18@163.com}

\author{Yuan Wang}
\orcid{0009-0007-8604-6243}
\affiliation{%
 \institution{Central Conservatory of Music}
 \country{China}}
\email{22a056@mail.ccom.edu.cn}

\author{Xiaobing Li}
\orcid{0000-0003-0113-824X}
\affiliation{%
 \institution{Central Conservatory of Music}
 \country{China}}
\email{lxiaobing@ccom.edu.cn}

\author{Feng Yu}
\orcid{0009-0007-0607-7315}
\affiliation{%
 \institution{Central Conservatory of Music}
 \country{China}}
\email{yufengAI@ccom.edu.cn}

\author{Tao Yu}
\orcid{0000-0002-3818-5069}
\affiliation{%
 \institution{Tsinghua University}
 \country{China}}
\email{ytrock@mail.tsinghua.edu.cn}
\authornote{Corresponding Author}

\author{Qionghai Dai}
\orcid{0000-0001-7043-3061}
\affiliation{%
 \institution{Tsinghua University}
 \country{China}}
\email{qhdai@mail.tsinghua.edu.cn}
\authornotemark[2]

\renewcommand\shortauthors{Jin, Y., Qiu, Z., and Shi, Y. et al.}

\begin{abstract}
In this paper, we touch on the problem of markerless multi-modal human motion capture especially for string performance capture which involves inherently subtle hand-string contacts and intricate movements. 
To fulfill this goal, we first collect a dataset, named String Performance Dataset (SPD), featuring cello and violin performances. 
The dataset includes videos captured from up to 23 different views, audio signals, and detailed 3D motion annotations of the body, hands, instrument, and bow. 
Moreover, to acquire the detailed motion annotations, we propose an audio-guided multi-modal motion capture framework that explicitly incorporates hand-string contacts detected from the audio signals for solving detailed hand poses. 
This framework serves as a baseline for string performance capture in a completely markerless manner without imposing any external devices on performers, eliminating the potential of introducing distortion in such delicate movements. 
We argue that the movements of performers, particularly the sound-producing gestures, contain subtle information often elusive to visual methods but can be inferred and retrieved from audio cues. Consequently, we refine the vision-based motion capture results through our innovative audio-guided approach, simultaneously clarifying the contact relationship between the performer and the instrument, as deduced from the audio. We validate the proposed framework and conduct ablation studies to demonstrate its efficacy. Our results outperform current state-of-the-art vision-based algorithms, underscoring the feasibility of augmenting visual motion capture with audio modality. To the best of our knowledge, SPD is the first dataset for musical instrument performance, covering fine-grained hand motion details in a multi-modal, large-scale collection. 
It holds significant implications and guidance for string instrument pedagogy, animation, and virtual concerts, as well as for both musical performance analysis and generation. Our code and SPD dataset are available at \url{https://github.com/Yitongishere/string_performance}.
\end{abstract}

%
%


%
%


\maketitle

\section{Introduction}

\begin{table*}[ht]
\caption{A summary of commonly used music performance datasets, detailing their focused instruments, number of excerpts, duration, and forms of content.}
\centering
\begin{tabular}{p{5cm}p{2.3cm}p{0.8cm}p{1.2cm}p{2.1cm}p{4.3cm}}
\hline
\textbf{Dataset} & \textbf{Instrument} & \textbf{Pieces} & \textbf{Duration} & \textbf{Camera Views} & \textbf{Mocap Annotation} \\ \hline
\rowcolor{gray!10} \multicolumn{6}{c}{\textit{Marker / Sensor Based Dataset}} \\

TELMI \cite{volpe2017multimodal}   & Violin   & 41   & 2.4 h   & 3 + 13 (infrared) & Body, Instrument, Bow  \\

QUARTET \cite{papiotis2016computational}  & String quartet  & 30  & 0.5 h  & 1 + 26 (infrared) & Body, Instrument, Bow\\ 

MMG \cite{perez2016estimation}  & Guitar  & 10  & 0.17 h  & N/A & Body, Hands, Instrument  \\

EEP \cite{marchini2014sense}  & String quartet  & 23  & N/A  & 0 (wired EMF) & Bow  \\ 

Bowstroke \cite{young2007bowstroke}  & Violin  & N/A  & N/A  & 1 & Bow \\

\hline \rowcolor{gray!10} \multicolumn{6}{c}{\textit{Markerless Dataset}} \\

CCOM-HuQin \cite{zhang2022ccom}  & HuQin  & N/A  & 1.29 h  & 3 & N/A \\ 

URMP \cite{li2018creating}    & Multi-instrument   & 44   & 1.3 h   & 1 & N/A  \\ 

C4S \cite{bazzica2017vision}  & Clarinet  & 54  & 4.5 h  & 1 & N/A  \\

ENST-Drums \cite{gillet2006enst}  & Drum kit  & N/A  & 3.75 h  & 2 & N/A \\ 

\textbf{SPD (ours)} & \textbf{Cello and Violin}  & \textbf{120}  & \textbf{3.0 h}  & \textbf{23} & \textbf{Body}, \textbf{Hands}, \textbf{Instrument}, \textbf{Bow}  \\ \hline

\end{tabular}

\label{tab:1}
\end{table*}

In the broad spectrum of human movement, the playing of instruments stands as an intricate demonstration of human fine motor skills. Thus, traditional video recording often fails to reveal such details. The challenge lies in the delicate dance of a musician's fingers, especially within the realm of string instruments, where the absence of fixed keys—characteristic of pianos or brass instruments—imparts a remarkable degree of freedom to the performer. While current motion capture (MoCap) methodologies excel in seizing common movements, they often stumble in accurately replicating these domain-specific actions, primarily due to a lack of specialized motion data. Despite these formidable obstacles, the capturing and subsequent recreation of string instrument performance holds vast potential in applications such as string instrument pedagogy, virtual concerts, and animation industries \cite{wheatland2015state}. These fields are currently underserved by existing technologies, implying significant development opportunities. It is worth noting that string performance is an art form that intertwines both visual and auditory modalities, engaging both senses to culminate in a comprehensive musical expression. The movements of performers serve as the physical foundation for the music creation, embodying an ambiguous but inevitable connection between action and sound. As a result, our key observation involves establishing and further leveraging the subtle correlations between the music conveyed through audio and movement depicted visually during string performance. This approach may facilitate breakthroughs in this domain-specific motion capture, transcending related solutions grounded solely on a single modality.

The capture and analysis of musical instrument performance movements have received increasing attention in recent years \cite{perez2019finger, papiotis2016computational, zhang2022ccom, mcpherson2022oxford}. Currently, a common challenge in this field is the lack of available data, primarily due to the prohibitively high cost associated with manual annotation. Table~\ref{tab:1} lists some relative data sources, however, some of them are limited in scale, resolution, and number of viewpoints or only provide raw video without extracting motion information. Naturally, MoCap becomes indispensable, enabling in-depth analysis and understanding beyond the scope of standard video footage. MoCap methods are typically classified into marker-based and markerless (or marker-free) approaches depending on whether markers are involved. At present, the majority of existing studies utilize marker-based approaches with devices such as infrared sensors, electromagnetic systems and inertial systems to capture motion for various musical instrument performances, including clarinets \cite{teixeira2015motion}, guitars \cite{perez2019finger}, drums \cite{gonzalez2019characterizing}, violins \cite{volpe2017multimodal, young2007bowstroke}, cellos \cite{gonzalez2019characterizing}, pianos \cite{tits2015feature, payeur2014human}, instrument duos \cite{jakubowski2017extracting, thompson2017interpersonal} and string quartets \cite{papiotis2016computational}. The primary advantage of marker-based methods lies in their precise positioning without the need for additional visual algorithms. However, they require special environmental setups such as infrared or magnetic fields, imposing certain demands on the experimental environment. Moreover, it is inevitable that markers deployed on the performer's torso, limbs, and hands introduce varying degrees of hindrances to the musical performance. This obstacle makes it impractical to place markers on every finger joint, thereby preventing performers from executing their performance smoothly \cite{perez2019finger}, resulting in the loss of capturing the most critical sound-producing finger movements. On the other end of the spectrum, the markerless approach does not impede performers and enables the investigation of as many points as desired in standard imagery, laying the necessary groundwork for exploring detailed finger movements. However, the markerless approach lacks guaranteed accuracy, and currently, advanced pose estimators are trained on generic data without validation for performance in specific scenarios like musical instrument playing, where complex human-object occlusion and interaction relationships are involved, indicating a need for improvement. As of now, there has been no previous work focused exclusively on exploring fine-grained instrument-playing motions using the markerless MoCap approach, suggesting potential opportunities ahead. Another noteworthy aspect is that regardless of the involvement of markers, both MoCap approaches struggle to capture the contact information between the performer's note-playing fingers and the instrument. However, this contact is crucial as it serves as a key signal for motion capture or motion generation during musical performance.

To address these challenges, we first captured a multi-view multi-modal dataset of string instrument performance on a totally marker-free basis, named String Performance Dataset (SPD). We publish the dataset and hold the belief that this non-intrusive setup is pivotal in enabling musicians to perform with freedom and naturalness, devoid of any movement distortion. This ensures that every subtle gesture they display holds inherent value, thereby enhancing the significance of our precise annotation method. In practice, we found that even with 23 cameras and the state-of-the-art MoCap networks, we still cannot obtain accurate hand poses due to the complex finger movements and the absence of constraints on hand-string interaction. Thus, we also present a baseline framework that models both the performer and instrument with a combination of audio signals to enhance the vision-based MoCap. The proposed framework finally yields pose annotations aligned with hand-string interactions, resulting in more reasonable and accurate MoCap outcomes, especially for subtle finger movements. To conclude, our main contributions are: 1) The first large-scale multi-modal MoCap dataset for string instrument performance with fine-grained instrument-sound-aligned hand poses. 2) An audio-guided multi-modal framework for enhancing markerless motion capture for string performance.

\section{Related Work}

\textbf{Instrument performance dataset:} Motion capture in musical instrument performance requires data of visual modality. However, visual resources are scarce compared to audio data. Table~\ref{tab:1} lists several datasets featuring visual modalities or MoCap data of musical instrument performance, along with their detailed information. The datasets on a marker-free basis like URMP \cite{li2018creating}, C4S \cite{bazzica2017vision}, ENST-Drums \cite{gillet2006enst} and CCOM-HuQin \cite{zhang2022ccom} merely contain raw videos from limited views, without capturing performance movements or providing related motion data. Most marker-based datasets only contain coarse-grained MoCap annotation without hand details \cite{papiotis2016computational, volpe2017multimodal, marchini2014sense, young2007bowstroke}, consequently missing the most essential movements in instrument performance scenarios. Although Multi-modal Guitar \cite{perez2016estimation} includes finger movements for guitar playing, the dataset is quite limited, containing only 10 data entries totaling 10 minutes. Additionally, \cite{simon2017hand} collects cello performance videos to validate their hand keypoint detection algorithm. However, this work does not specifically focus on instrument performance, and as a result, the amount of data is quite limited and fails to model the instrument.

\textbf{Marker-based instrument performance MoCap:} This type of approach utilizes sensors or markers attached to performers and instruments to capture their spatial location. Specific technologies include infrared markers \cite{roze2018assessing}, electromagnetic field (EMF) sensing \cite{papiotis2016computational}, and inertial measurement unit (IMU) \cite{young2007bowstroke}. Some scholars also integrate multiple sensors to improve MoCap accuracy. \cite{schoonderwaldt2009extraction} combine the infrared markers and IMU to capture motion for violin performance. \cite{gonzalez2019characterizing, volpe2017multimodal} introduce the electromyogram (EMG) as an auxiliary tool to assist in instrumental performance MoCap. Among these technologies, the infrared method is most popular owing to its high capturing accuracy with relatively small marker sizes (3 mm around), which has also been extended to multiple musicians like violin duets \cite{thompson2017interpersonal}, string quartets \cite{maestre2017enriched}, and musical ensembles \cite{hilt2019multi}. However, the attached marker inevitably disrupts the hand or finger movements of the performer. \cite{perez2016estimation, perez2019finger} we mentioned above, investigate the finger-string interaction in guitar playing, placing markers on the dorsal side of each finger articulation. However, the contact between fingers and strings primarily occurs on the palm side. This deviation, compounded by errors stemming from marker size, becomes substantial when analyzing intricate finger movements, rendering this approach both intrusive to the playing and unreliable regarding accuracy. Therefore, most of the above works only focus on larger body parts like arms and torso to explore expressive movements, failing to effectively capture the sound-producing actions of the performer's hands. 

\textbf{Vision-based instrument performance MoCap:} Vision-based motion capture technology facilitates the analysis of myriads of points from videos, which is not constrained by the number of markers. This provides a necessary foundation for exploring the intricate finger movements of musicians. \cite{hadjakos2013motion} apply a purely vision-based approach to analyze violin performances with Kinect. However, they focus mainly on the performer's head movements and do not delve into finger movements. Beyond research specifically targeting instruments, some general human pose estimators can also capture these scenarios with vision-based algorithms \cite{cao2017realtime, yang2023effective, lugaresi2019mediapipe, zhang2020mediapipe}. Yet, they fall short in training with domain-specific data, underscoring the necessity to improve their accuracy in detecting the unique postures typical in musical performances.

\textbf{Multi-modal approach in musical performance:} Recent works integrating audio modality focus on generating movements for violin playing \cite{hirata2022audio, hirata2021bowing, kao2020temporally, shrestha2022aimusicguru} and singing \cite{pan2022vocal}. Earlier studies on hand and finger modeling and animating explore mapping sheet music (or symbolic music notation) to finger movements using rule-based approaches, involving animation generation for guitar \cite{elkoura2003handrix}, violin \cite{kim2000neural}, and piano \cite{zhu2013system, kugimoto2009cg}. However, these vague mappings from audio or sheet music to performance are unable to precisely replicate the real-world actions, resulting in relatively rough playing motions that fail to capture the nuanced original movements.

\section{Methodology}
\begin{figure}[t]
    \centering
    \includegraphics[width=\the\columnwidth]{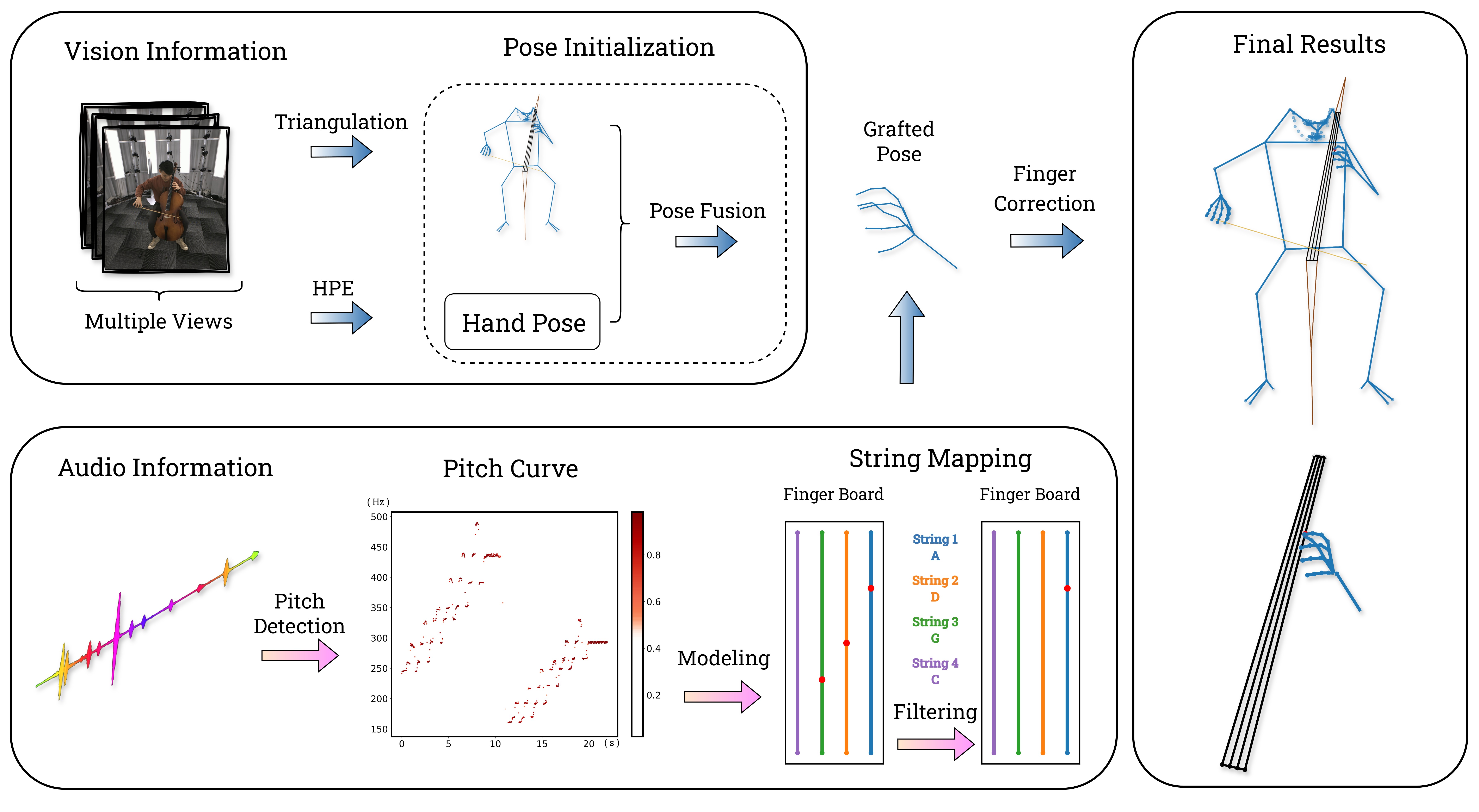}
    \caption
        {
            The pipeline combines the information extracted from the visual and auditory inputs. The following sections elaborate on detailed explanations of each part.
            \label{flowchart}
        }
\end{figure}

In this section, we present the process of constructing the String Performance Dataset by providing an in-depth look at the complete pipeline of our MoCap framework, which is specially designed to capture string instrument performance. This includes detailed coverage of all stages from data acquisition to final output. As illustrated in Fig.\ref{flowchart}, under the premise of not involving markers or sensors, we integrate domain knowledge of the specific string instruments (cello or violin) and features extracted from the music produced by the instrument playing to constrain and further optimize the reconstruction results achieved solely through vision-based methods. We concentrate on the cello and violin, two of the most mainstream classical string instruments, which exhibit significant differences in playing postures, instrument sizes, and musical ranges.

\subsection{Data Acquisition}
\begin{figure}[h]
    \centering
    \includegraphics[width=\the\columnwidth]{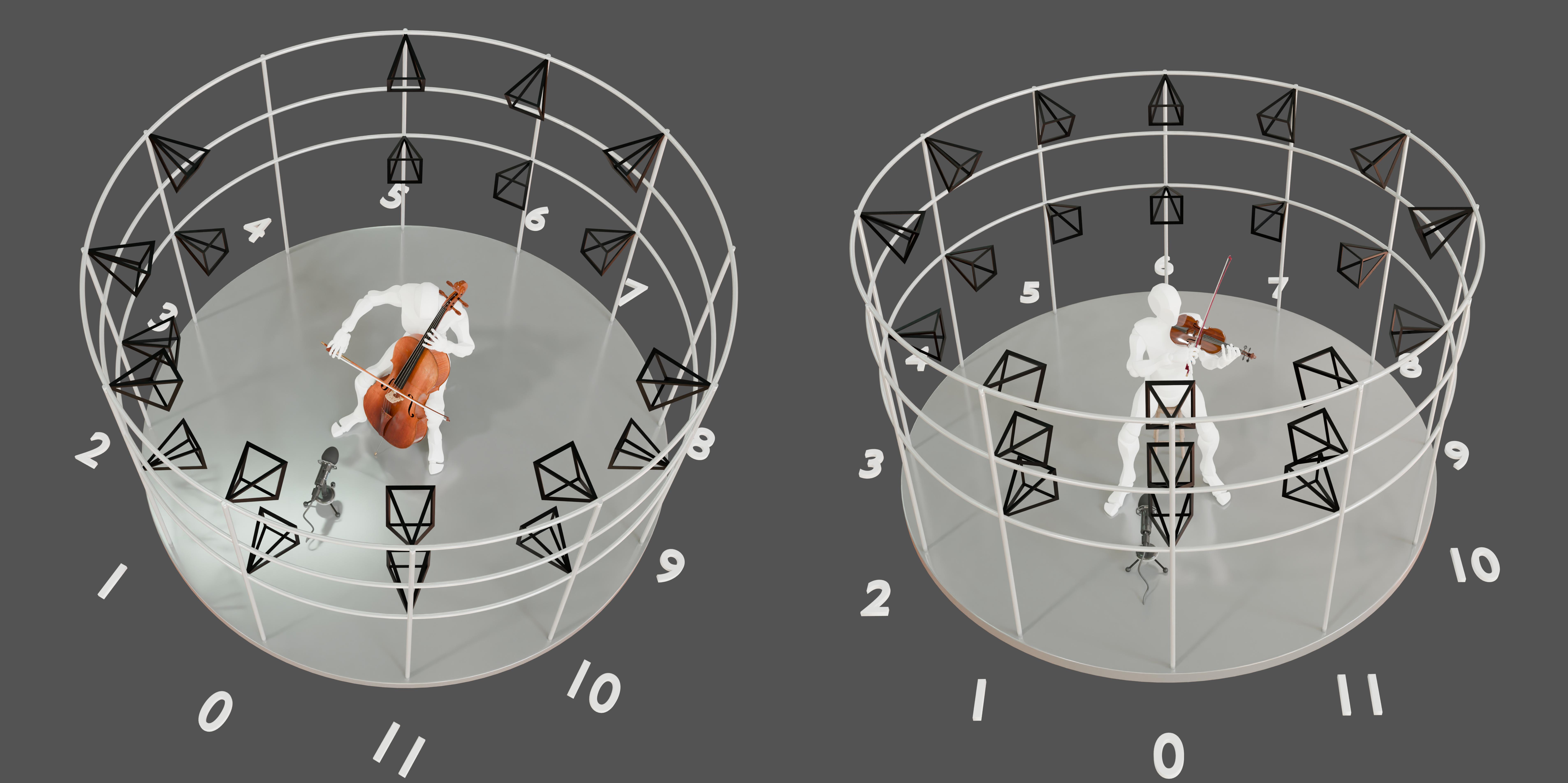}
    \caption
        {
            Our recording setup, with slight differences between the cello and violin scenarios in the camera numbers and positions. 20 cameras for the cello and 23 cameras for the violin.
            \label{cam}
        }
\end{figure}

MoCap for the string performance suffers from severe occlusion caused by the complex interactions between the performer and the instrument. To minimize this challenge, we establish a multi-view MoCap system, as shown in Fig.\ref{cam}. The system consists of up to 23 cameras distributed across 12 fixed poles positioned at the "hour marks" around the performer, covering a cylindrical space with a diameter of approximately 4 meters. While our system can function effectively in both cello and violin scenarios with identical camera settings, we adopt slightly different camera distributions to achieve a more comprehensive Line-of-Sight coverage by positioning additional cameras in front of the violin performers. This arrangement addresses challenges inherent in capturing violin performances, including variations in fingerboard orientations and the smaller size of the target compared to the cello. Each camera is focused on the performer at the center of the venue, and solid-colored curtains are placed around the venue to eliminate any potential distractions from the background. All used cameras are FLIR's ORX-10G-245S8C, featuring a resolution of 2656x2300, a frame rate of 30 fps, and input/output synchronization. For audio recording, we use a Sony ICD-PX470 microphone fixed in front of the performer. In addition, a manual clap is applied to align the video and audio signals before each performance recording. 

The data we collected includes performances from a total of 9 cellists and violinists, ranging from amateurs and conservatory students to acclaimed professionals. We apply the standard of 440 Hz for the A4 note during string tuning. The repertoire spans a diverse range, including scales, études, classical compositions, and modern pop music, encompassing 120 pieces with a cumulative duration exceeding 3 hours. This breadth ensures a robust and versatile dataset, suitable for a wide range of applications in musical research and practice. Most importantly, all performances take place in a completely marker-free setting, allowing musicians to fully express their movements. This greatly enhances the value of our subsequent detailed and nuanced capture of these actions.

\subsection{Pose Initialization}
In this subsection, we describe how we initialize the spatial information of the performer and instrument. Spatial information requires reconstruction from multiple views, hence the necessity to first obtain 2D-keypoint locations. Given that our points of interest are distributed across three parts: the human, the instrument, and the bow, each having distinct motion patterns, we employ different algorithms to extract them separately based on 2D images.

\textbf{Performer pose initialization:} Identifying body or hand movements is invariably complex, yet thankfully, a wealth of advanced methods, devices, and datasets \cite{zheng2023deep, li2019survey, wheatland2015state} exist to support our endeavors. We apply the open-source \textit{DWPose} model \cite{yang2023effective}, a state-of-the-art whole-body pose estimator that surpasses two widely used models OpenPose \cite{cao2017realtime} and MediaPipe \cite{lugaresi2019mediapipe, zhang2020mediapipe}, to identify keypoints of the performer from 2D images, including fingers, limbs, torso, and facial expressions, totaling 133 keypoints. DWPose model achieves precise localization in human keypoint detection, however, it may cause hand distortions that do not follow anthropometrics. Some extreme relationships of finger positions are unusual in daily life but frequently occur on the left hand of string instrument performers when conducting complicated playing techniques. To tackle this issue, we specifically train a hand pose estimator (HPE) using the combination of several commonly used datasets for hand pose estimation, including  InterHand2.6M \cite{moon2020interhand2}, DARTset \cite{gao2022dart}, BlurHand \cite{oh2023recovering}, and HanCo \cite{zimmermann2021contrastive}. In the HPE implementation, we follow the backbone architecture and implementation details of InterNet which is outlined in \cite{moon2020interhand2}. However, we make a notable modification to its output layer. Instead of outputting hand joint positions, the final layer is transitioned to predict a 6D representation for 3D hand joint rotation \cite{zhou2019continuity}, constrained by the MANO model \cite{romero2022embodied}. This modification yields three pivotal effects. First, integrating the MANO model introduces anthropometrics priors of the human hand, thereby guiding the HPE to emphasize the correctness and reasonableness of hand poses. Secondly, the prediction of 3D hand joint rotation provides a complete kinematic chain from the wrist joint to each fingertip, bringing forth the feasibility of further hand pose correction based on inverse kinematics methods, as elaborated later. Lastly, supported by comprehensive empirical results in \cite{zhou2019continuity}, the continuous 6D representation has been proven to be beneficial for training deep neural networks, leading to faster convergence and lower loss when compared to commonly used but discontinuous ones, such as quaternions and Euler angles. As for the inference process of performers' hand poses, we conduct predictions of the HPE model on various views of imagery. Subsequently, in order to achieve the integration of the results from multiple perspectives, the inferred 6D representations are converted into quaternions, following which they will undergo interpolation based on the assigned weights of each camera. We assign weights according to the relative orientation between the camera and the performer, prioritizing perspectives that offer clearer views of the left hand. Up to this point, we obtain the performer's left-hand pose prediction by the HPE model across multiple viewpoints. However, the question arises: how do we "graft" such hand pose prediction onto the predictions of the rest of the body parts from DWPose to fuse the results? The key lies in determining the angle at which the left wrist joint connects. First, we align the wrist positions from both results. Then we adopt the approach of inverse kinematics, taking the hand pose prediction from DWPose as the target and continuously rotating and translating the whole hand pose prediction from the HPE model around the wrist joint until getting the minimal Euclidean distance of all corresponding hand joints between the two sets of results. This optimization process can be expressed as:
\begin{equation}
\label{eqn:01}
(\theta_{\text{w}}^*,\text{T}^*) = \underset{\theta_{\text{w}}, \text{T}}{\text{argmin}} \, \left( \sum \|J_{\text{HPE}}(\theta_{\text{w}}, \text{T}) - J_{\text{DWPose}}\| \right),
\end{equation}
where \(J_{\text{HPE}}\) and \(J_{\text{DWPose}}\) represent the positions of all hand keypoints predicted by the HPE and DWPose respectively. \(\theta_{w}\) and \(T\) denote the rotation matrix at the wrist joint and the translation vector to be optimized from the HPE prediction. As for the rest of the body parts beyond the left hand, we use the results predicted by the DWPose model directly.

\begin{figure}[t]
    \centering
    \includegraphics[width=\the\columnwidth]{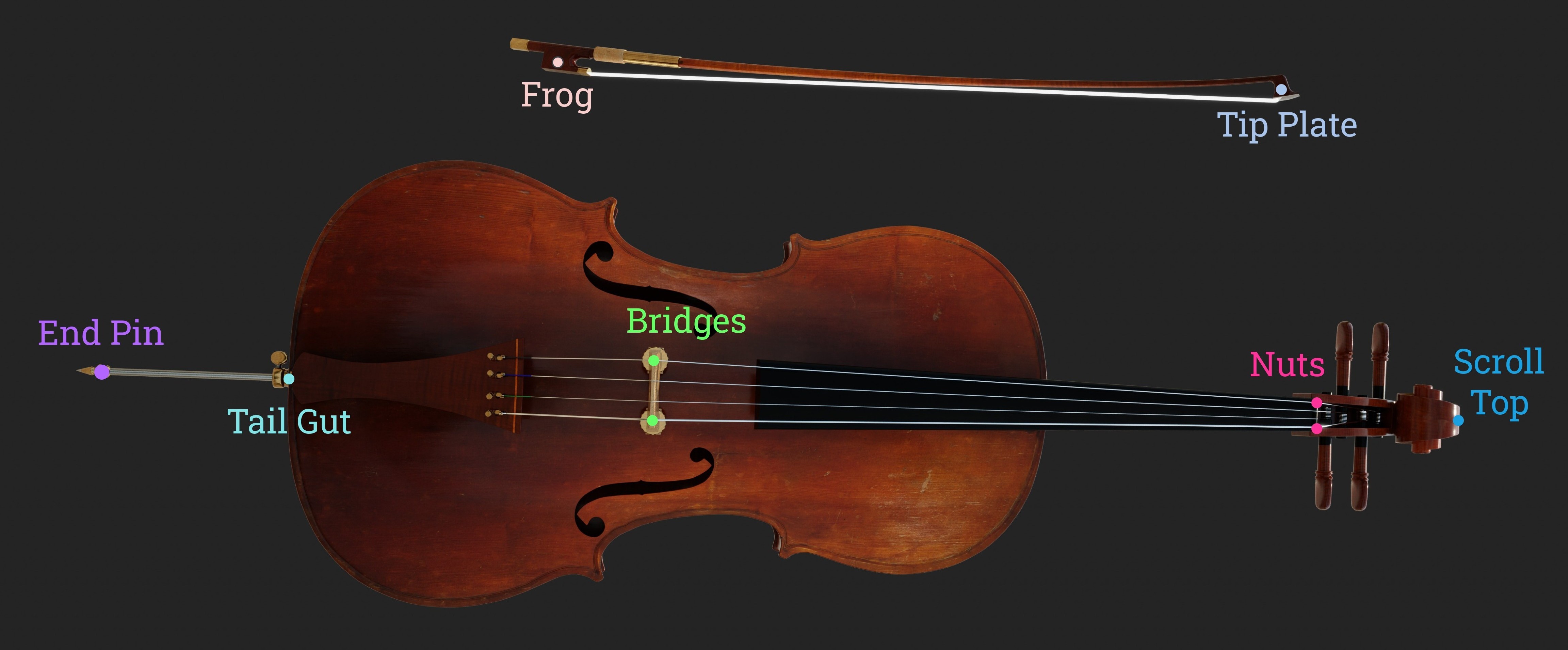}
    \caption
        {
            Illustration of cello keypoints.
            \label{tap}
        }
\end{figure}

\textbf{Instrument pose initialization:} Modeling the instrument is essential, particularly the strings, as they are the most direct interactive elements between the performer and the instrument, encompassing key sound-producing finger movements. Currently, there is no established ready-to-use method for localizing and modeling string instruments. However, due to the rigid structure of the instrument, we can infer its overall state by determining a few keypoints after taking its geometric structure as prior knowledge. We leverage Google's TAPNet \cite{doersch2023tapir} for keypoints tracking as it exhibits exceptional performance in tracking such gentle and slight rotation or displacement of the instrument body during the performance. 
We specify the points of our interest in the first frame of the video being analyzed, and the corresponding positions of these points will be automatically retrieved in the subsequent frames. In Fig.\ref{tap}, we illustrate how we select keypoints to be tracked on the cello. A similar setup is used for the violin, except for removing the end pin and the tail gut. Once the nut and bridge are located, the positions of the strings can be deduced based on geometric relationships. Direct positioning is avoided because the strings are less visible in the imagery.

\textbf{The bow:} The pose of the bowing hand (right hand) does not vary much when performing, but the speed, angle, and relative position between the bow and strings are viewed as key elements of the playing action. The mentioned TAPNet is not robust for bow tracking due to the distinct characteristics of bow movements compared to those of the instrument body. Bow movements involve extensive and high-speed movements, especially when playing intense notes. We trained our bow detection model based on YOLOv8 architecture \cite{YOLOv8} to identify the frog and tip plate of the bow. 

Next, we integrate keypoints from synchronized frames across various views through triangulation, obtaining the three-dimensional spatial information of the scenario. In this process, no single camera can independently capture all keypoints of interest due to occlusions caused by the instrument and the performer, self-occlusion by the performer, and even the obstructive relationship between the bow and the instrument. To address this issue, we apply the Random Sample Consensus algorithm (RANSAC) to exclude outliers in the reconstruction of each spatial point. 

All the motion capture processes described above are completed on a purely visual and marker-free basis, achieving the reconstruction of keypoints in three-dimensional space for the performer, the instrument, and the bow in string instrument performance scenes. In the following subsection, we elaborate on how we use the audio-guided approach to further optimize the results.

\subsection{Audio-guided Multi-Modal Motion Capture}
In string instrument performance, playing movements serve as the physical basis of music production. Building on this idea, we believe that music itself contains key information to infer the interaction between the fingers and the instrument. By utilizing our designed Pitch-Finger model and incorporating domain knowledge from string instruments, we map the musical pitch information extracted from audio onto specific locations on the fingerboard of the instrument. This provides the foundation for correcting our initial motion results obtained from the visual modality. 

\subsubsection{Pitch Detection}
Unlike noise or human speech, string instruments produce sound with a regular and repetitive waveform, resulting in a distinct sound frequency, i.e., the pitch. The performers alter the pitch by adjusting their left-hand position on the fingerboard and the strings they touch. Meanwhile, the mapping between the playing movements and pitch is much clearer than that of the other sound elements such as timbre, duration, and intensity. We utilize the CREPE pitch tracker \cite{kim2018crepe}, which achieves up to 99.9 \% accuracy with a threshold of 25 cents in monophonic pitch detection, to determine the pitch curve of our recorded audio. In this process, we maintain consistency between the frequency of applying pitch estimation in music and the frame rate of our video recording (30 fps), ensuring a one-to-one correspondence between each frame and its associated pitch value.

\subsubsection{Pitch-Finger Model}

\begin{figure}[t]
    \centering
    \includegraphics[width=\the\columnwidth]{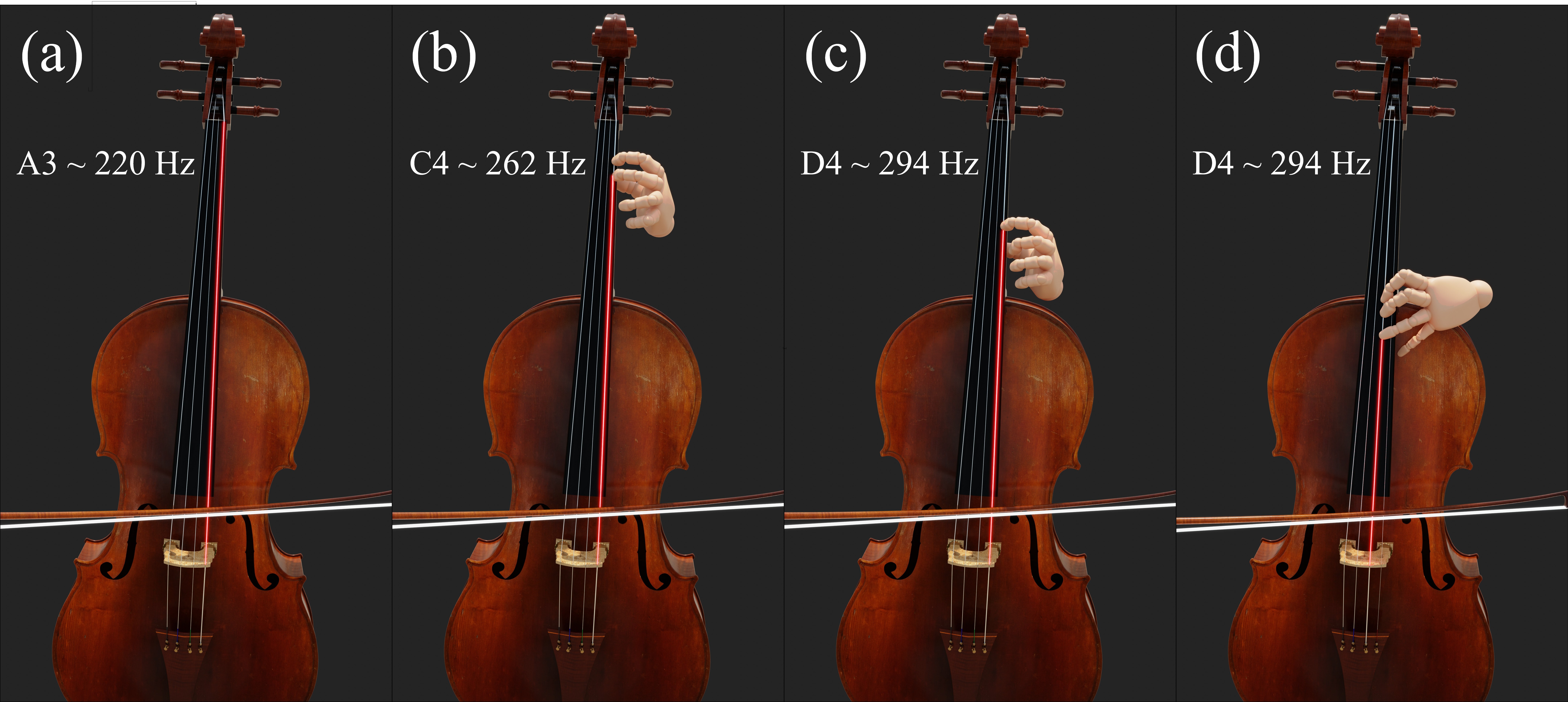}
    \caption
        {
            Using the cello as a reference, we present examples of various note-playing finger positions alongside their corresponding vibrating lengths (highlighted in red) and pitch values. \textbf{(a)} represents an open string note, producing the lowest pitch achievable on the excited string. \textbf{(b)} and \textbf{(c)} demonstrate different finger positions on the same string, resulting in different pitches. \textbf{(c)} and \textbf{(d)} illustrate instances in which varying finger positions on different strings produce the same pitch.
            \label{vibrating_length}
        }
\end{figure}

The vibrating length refers to the portion of the string that actually vibrates to produce sound, highlighted in red in Fig.\ref{vibrating_length}, indicating the pitch. The relationship between pitch \(P\) and vibrating length \(L_{vib}\) can be presented as:
\begin{equation}
\label{eqn:02}
P = \frac{F \cdot L_{fund}}{L_{vib}},
\end{equation}
where \(L_{fund}\) is the full length of the string, known as fundamental length. \(F\) represents the frequency of vibration along the fundamental length (Fig. \ref{vibrating_length}-\textit{a}). When a string is excited by the bow, the lowest sound it can produce is its fundamental frequency, while theoretically, there is no upper limit for the highest pitch. Therefore, given a note whose pitch is above the fundamental frequency of a certain string on a string instrument, the vibrating length of the target string can be uniquely determined by Equation~\eqref{eqn:02}. 

Performers use their left hands to adjust the vibrating length by changing the note-playing finger position, at which the string is pressed (Fig. \ref{vibrating_length}-\textit{b} and \textit{c}). With the detected pitch value, we aim to determine the expected note-playing finger position deduced from the audio, referred to as the audio-guided position in the following discussion. The audio-guided position serves as the target for correcting finger movements. However, string instruments typically have four strings, with some overlap in pitch range. This overlap can lead to ambiguous finger placement, as the same pitch might be produced by multiple finger positions on different strings (demonstrated in examples \textit{c} and \textit{d} of Fig. \ref{vibrating_length}). Therefore, accurately identifying the audio-guided position solely based on pitch poses a significant challenge. To determine it precisely, we use our initial results from the vision-based algorithm for the left hand as guidance. Initially, we identify multiple potential finger positions based on the detected pitch, all of which are reasonable for producing the target note frequency. Then, we ascertain the actual finger position of the performers by referring to their wrist position from the results obtained in section 3.2. The feasibility of this step is twofold. On the one hand, as illustrated in Fig. \ref{vibrating_length} -\textit{c} and \textit{d}, the apparent difference in finger position of the same note on different strings greatly reduces the threshold of filtering the final audio-guided position from potential ones. On the other hand, benefiting from the well-adjusted multi-view setup, our previous results based solely on visual detection are quite reliable for further optimization according to the audio signals. Finally, once the audio-guided position is determined, we bind it to its nearest fingertip, forming a pair between the target and the fingertip to be corrected. At each time step, we focus on adjusting the note-playing finger to the corresponding time.

In addition, we incorporate domain knowledge of string instruments to guide the algorithm in binding the audio-guided position to the note-playing finger. When performing the vibrato effect, the pitch exhibits rapid and slight fluctuations, and the note-playing finger for these consecutive frames remains unchanged. The physical premise of this vibrato effect is the fast overall trembling of the left hand. However, this trembling can change the distance between the audio-guided position and each fingertip, affecting the binding relationship. To avoid such circumstances, we assess the pitch curve by detecting the pattern of pitch variation. If the pitch variation across consecutive frames is within ±30 cents (100 cents per semitone), vibrato is confirmed, and the binding relationship from the onset to the end of the vibrato remains constant.

Identifying the binding relationship is crucial for further detailed optimization of the note-playing finger. The external force exerted by the fingerboard on the left hand during string instrument performance can cause varying degrees of joint deformation. Such deformation induces deviation in the prediction as our used hand-pose estimator (HPE) is trained on data where the hand is not subjected to external forces. Hence, we further employ inverse kinematics to guide the note-playing fingertip towards the audio-guided position, which can be expressed as:
\begin{equation}
\label{eqn:03}
(\theta_{\text{d}}^*, \theta_{\text{p}}^*) = \underset{\theta_{\text{d}}, \theta_{\text{p}}}{\mathrm{argmin}} \, \|J_{\text{tip}}(\theta_{\text{d}}, \theta_{\text{p}}) - J_{\text{audio-guided}}\|.
\end{equation}
We take the distance between the audio-guided position (\(J_{\text{audio-guided}}\)) and the tip of the note-playing finger (\(J_{\text{tip}}\)) as the cost function. We then fine-tune the rotation angles (\(\theta_{\text{d}}, \theta_{\text{p}}\)) of the Proximal Interphalangeal joint and the Distal Interphalangeal joint of the note-playing finger using the L-BFGS-B algorithm \cite{zhu1997algorithm}, a method previously employed in \cite{tsang2005helping}, affirming its efficiency in rectifying hand poses. This ensures optimal alignment between the note-playing fingertip and the audio-guided position. Even though there might theoretically be multiple solutions for the target finger angles, the strong constraints of audio-guided position and the proximity of our initial results to these optima enable us to consistently achieve desirable finger postures within a limited solution space. Additionally, employing shorter optimization iteration step size helps avoid unreasonable solutions that adhere only to mathematics but not to anthropometrics.

\section{Results and Comparison}
\begin{figure}[t]
    \centering
    \includegraphics[width=\the\columnwidth]{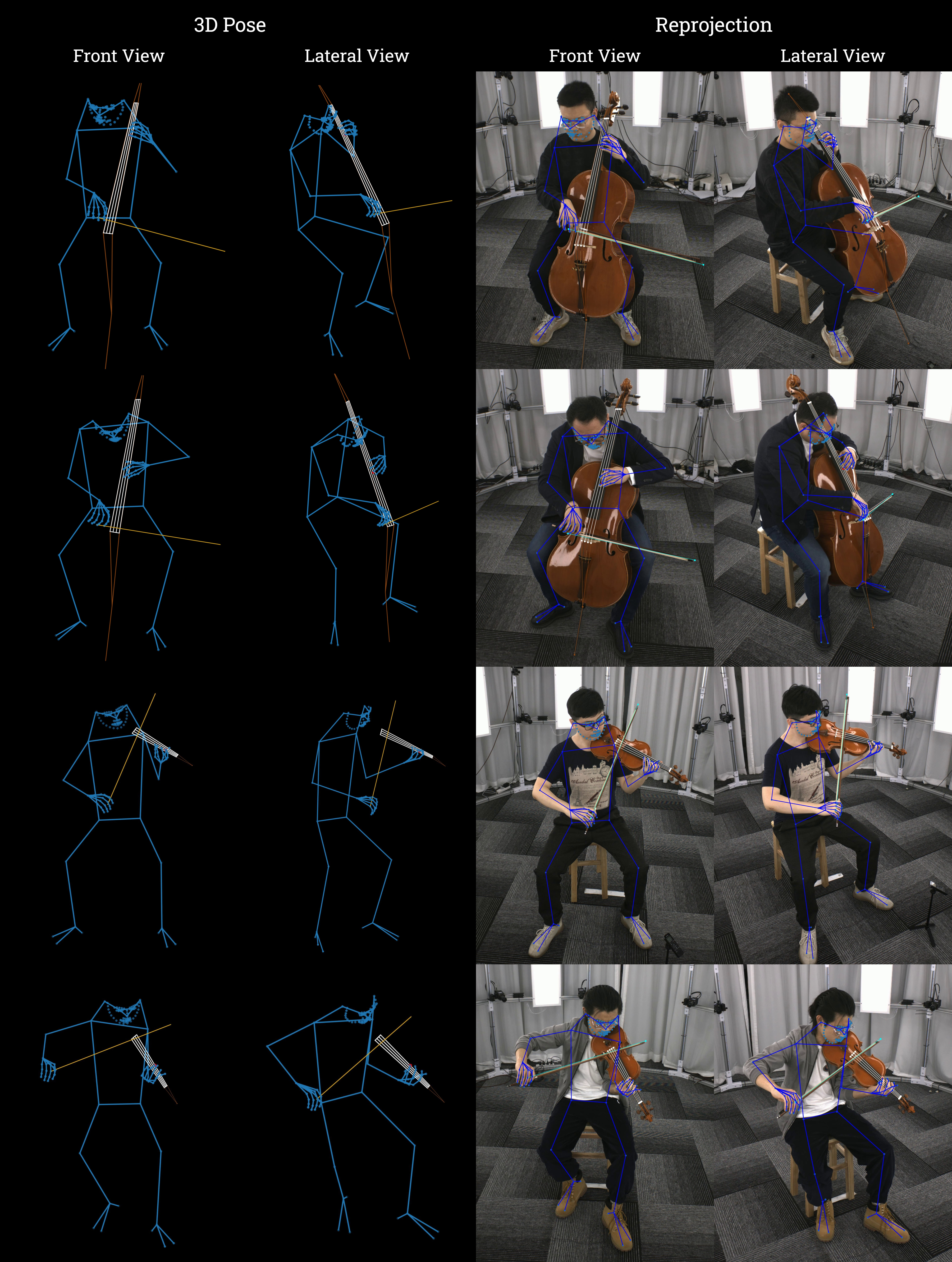}
    \caption
        {
            Demonstration of our results through reprojection and 3D visualization from different views.
            \label{3d_overlay}
        }
\end{figure}

With our proposed multi-modal MoCap framework, 3D movement information can be automatically extracted from the original multi-view videos. This 3D representation not only includes motion capture data of the performer and instrument but also accurately replicates the performer's hand contact and relative position with the instrument. In Fig.\ref{3d_overlay}, we present the global 3D results observed from multiple views and validate the reprojection results by overlaying the original images. Our 3D performance motion reconstruction demonstrates reasonable results for both cello and violin performance. 

\begin{figure}[ht]
    \centering
    \includegraphics[width=\the\columnwidth]{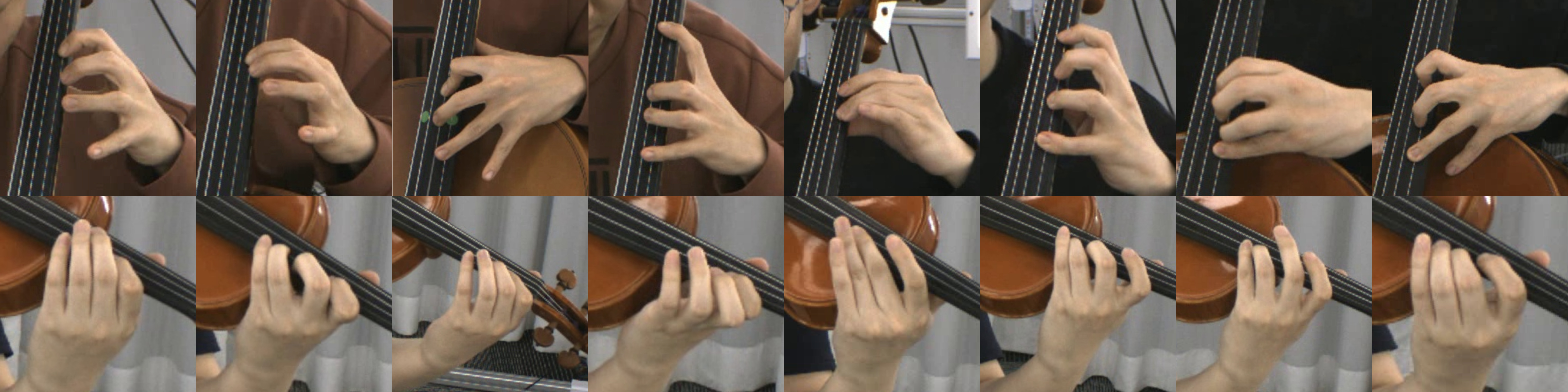}
    \caption
        {
            Some examples from the test set are presented, with a frontal zoomed-in view of the left hand, covering various holds. The test set contains 200 frames, each of which corresponds to 4 images of different viewpoints. We manually annotate these images before the triangulation to obtain the ground-truth 3D annotations of corresponding frames.
            \label{test_set}
        }
\end{figure}

\begin{table*}[h]
\caption{Ablation study: Evaluation on MPJPE and contact deviation, along with the improvements over the baseline (DWPose) results.}
\centering
\begin{tabular}{p{3.6cm}|rp{2cm}lp{1.5cm}rp{2cm}lp{1.5cm}rp{2cm}lp{1.5cm}}
\hline

Method & \multicolumn{2}{c}{MPJPE (whole hand)}\hspace{1cm} & \multicolumn{2}{c}{MPJPE (note-playing finger)}\hspace{1cm} & \multicolumn{2}{c}{Contact Deviation}\hspace{1cm}  \\ \hline

DWPose                        &      17.00  & \hspace{0.1cm} ---      &   16.14 & \hspace{0.1cm} ---       &     22.40 & \hspace{0.1cm} ---  \\

HPE (w/o Audio-guided)   &      15.27   & \hspace{0.1cm} $\downarrow$ 10.2 \%   &    14.95 &  \hspace{0.1cm} $\downarrow$ 7.4 \%   &     16.06 & \hspace{0.1cm} $\downarrow$ 28.3 \% \\ 

HPE + Audio-guided            &      \textbf{14.93}   & \hspace{0.1cm} $\downarrow$ 12.2 \%   &    \textbf{13.27}  & \hspace{0.1cm} $\downarrow$ 17.8 \%   &     \textbf{5.19} & \hspace{0.1cm} $\downarrow$ 76.8 \% \\ \hline

\end{tabular}

\label{tab:2}
\end{table*}

\begin{figure*}[t]
    \centering
    \includegraphics[width=\textwidth]{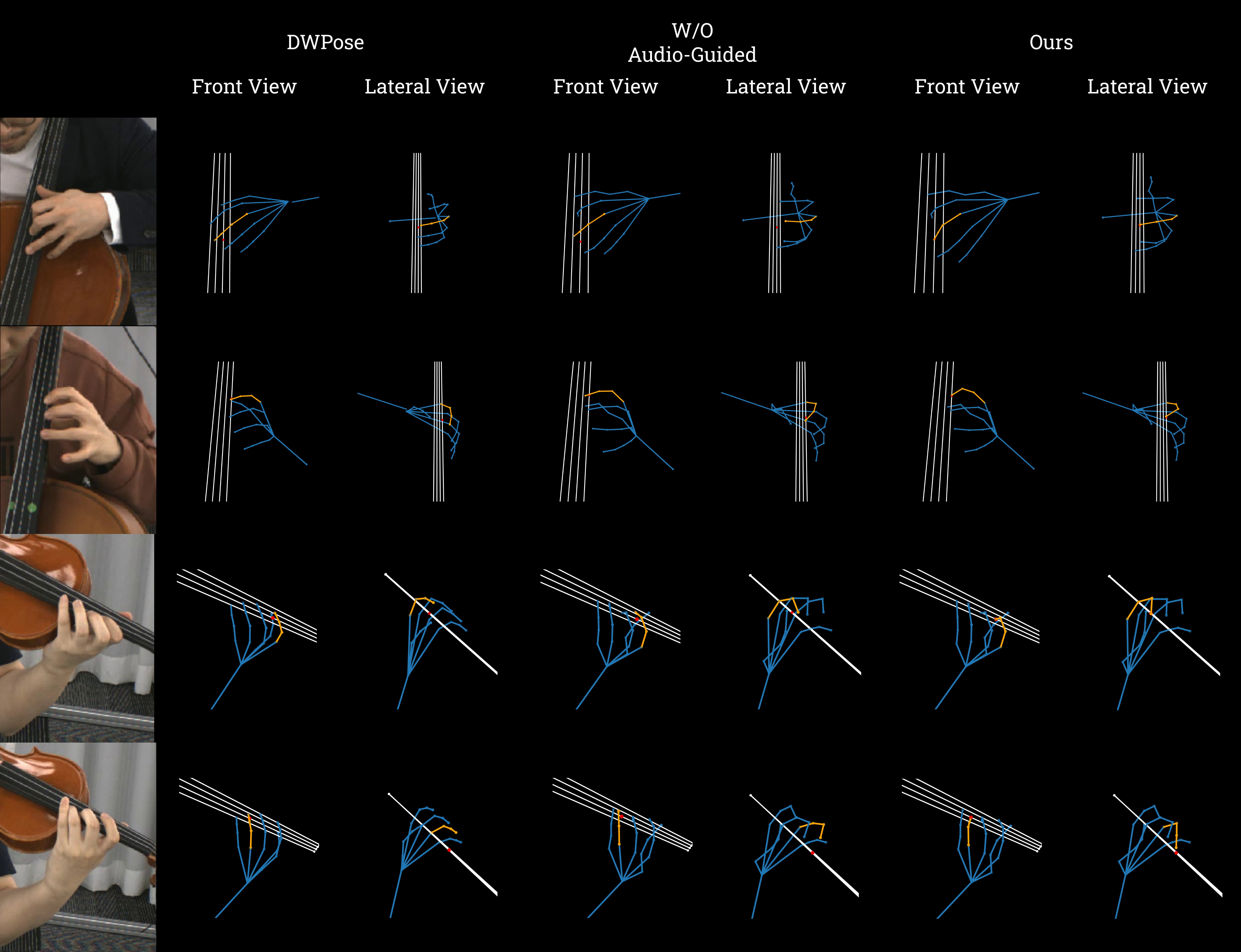}
    \caption
        {
            Close-up details show a comparison of various hand positions during performance with different degrees of optimization. Column "DWPose" presents the results from the state-of-the-art vision-based model without applying either our HPE model or the audio-guided approach, while the middle column shows the results from the HPE model only. The final column, labeled "Ours", incorporates our full approach.
            \label{ablation}
        }
\end{figure*}

Moreover, we conduct an ablation study to gain a more comprehensive understanding of how our audio-guided approach enhances vision-based MoCap results. 
For quantitative analysis, we utilize mean per joint position error (MPJPE) as our metric to assess both the entire left hand and specifically the note-playing finger. 
The MPJPE metric measures Euclidean distance (in millimeters) between estimated coordinates and the ground-truth coordinates after aligning their root joints (i.e., wrist joints) through translation.
While acquiring ground-truth 3D annotations is not straightforward since complex hand poses and occlusions are ubiquitous in string performance. 
To obtain ground-truth 3D annotations of each frame, we manually annotate the performer's left hand on 2D images from four different perspectives (simultaneously captured, with a clear presentation of the left hand), before triangulating these 2D annotations to 3D skeletons for metric calculation. 
All the annotators possess certain professional experience in playing string instruments to guarantee a reliable level of accuracy in annotating. 

A total of 200 frames are involved in the evaluation, covering comprehensive holds (wrist positions or finger placements with various hand poses on the fingerboard to produce specific notes) for both cello and violin performances. Some examples from our test set are listed in Fig.\ref{test_set}. Table~\ref{tab:2} summarizes the MPJPE results obtained from the DWPose model (the state-of-the-art vision-based model), our HPE model but without the audio-guided module, and our full approach respectively. 
In both the evaluation in terms of the entire hand and specifically the note-playing finger, our HPE model outperforms DWPose, and moreover, the integration of the audio-guided approach further enhances performance. Given that the note-playing finger is pivotal in sound production for string instrument performance, we conduct an analysis focusing exclusively on this crucial element, where the audio-guided approach exhibits a more pronounced effect.

Apart from MPJPE evaluation, we calculate the contact deviation to determine whether the contact relationship between the performer and the instrument is accurately reconstructed. This deviation is the Euclidean distance between the estimated position of the note-playing fingertip and the theoretical finger position inferred from the pitch. Before evaluation, these theoretical positions representing the expected locations of the note-playing finger's tip were manually verified. This measurement is presented in the last column of Table~\ref{tab:2}. 
The evaluation from this perspective indicates that integrating the audio-guided approach greatly improves the details of finger placement for pressing strings, thereby clarifying the contact relationship between the note-playing fingers and the instrument during the performance.

To further provide qualitative insights, we present detailed close-up views of the performer's left hand from various perspectives in Fig.\ref{ablation}, which corroborate the findings of the quantitative evaluation.

\section{Discussion}
The proposed motion capture framework and the String Performance Dataset have limitations that suggest future work. Unlike harmonic instruments like pianos, strings, as melodic instruments, mainly employ a monophonic playing style. However, polyphonic performance, where multiple strings are triggered simultaneously, may occur in certain pieces. The accuracy of polyphonic pitch detection poses a limitation of our currently employed approach, therefore, we do not include such a scenario as unreliable detection may mislabel note-playing fingers. In addition, we primarily focus on using the audio-guided approach to assist in capturing the performer's left hand. The potential for expanding the application of the audio-guided motion capture to include other body parts is on the horizon, necessitating the acquisition of additional constraints from the musical aspect. Apart from pitch information, other elements inherent in the audio, such as volume and note duration, remain unexplored. This exploration holds the promise of providing us with further insights for optimizing the results of the bow-holding hand. Furthermore, the recording frame rate is set at 30 frames per second, potentially posing a limitation for the upper bound of the fast motion.

Our contribution is capable of supporting research in string performance analysis and string instrument pedagogy. Future efforts aiming at achieving comprehensive coverage of string instruments could include the exploration of the viola and double bass, both of which share significant similarities with the strings family. Besides, with the availability of SPD, there is potential to create mappings between music and playing movements of string instruments. This advancement could facilitate the achievement of high-fidelity and professional-grade motion generation in animations or virtual avatars. On the other hand, in pursuit of flexibility in utility and a lightweight data collection system, the research could extend to monocular MoCap (from online videos or real-life performances recorded on mobile devices), which presents super challenges, especially with subtle instrumental performances. Nevertheless, we believe our audio-guided method demonstrates a promising and practical solution. It can also be combined with other newly proposed MoCap technologies via VR hardware \cite{han2022umetrack}, RF-vision \cite{zhang2023ochid}, MEMS-ultrasonic sensors \cite{zhang2023hand}, and even special patterns attached to the human body or hand \cite{chen2021capturing}. We leave this for our forthcoming research.

A broader perspective suggests that the application of the audio-guided approach can be extended to fields like sports (where the sound of ball impacts is linked with striking actions), dance performance (with movement transitions synchronized with musical cues), and other domains where human movements are closely associated with sounds. This could break through the inherent Line-of-Sight subjection in conventional camera-based MoCap to a certain extent, effectively improving results in situations with occlusion or contact.

\section{Conclusion} In summary, this paper presents the String Performance Dataset (SPD), which contains 3 hours of multi-view videos, corresponding 3D MoCap annotations, and synchronized audio signals. We also propose an audio-guided multi-modal framework to capture the fine-grained finger movements and instrument poses on a completely marker-free basis. It leverages the definitive correlations between audio and hand position, resulting in precise hand pose estimation that matches the exact instrument's sound. Such a novel audio-guided approach can serve as an inspiration for using audio information to enhance visual restoration in other scenarios. To this end, SPD is the first large-scale multi-modal MoCap dataset for string instrument performance with precise instrument-sound-aligned hand poses. The SPD surpasses existing similar datasets in terms of data volume, shooting angle variety, and the granularity of captured movement, effectively mitigating the data scarcity in the field of instrument performance. Based on the current progress, research on more generalized MoCap settings and the task of string performance generation will be carried out in the near future.

\begin{acks}
We thank Haotian Zhou, Xinghong Wang, Ziyi Huang, Shihao Yao, Yutong Ding, and Yuetonghui Xu for their performances in data acquisition. This work was supported in part by the National Key R\&D Program of China (No.2022YFF0902204), in part by the National Natural Science Foundation of China under Grant No.62171255, in part by the Tsinghua University-Joint research and development project under Grant R24119F0 JCLFT-Phase 1, in part by the "Light Field Generic Technology Platform" (Z23111000290000) of Beijing Municipal Science and Technology Commission, in part by the Guoqiang Institute of Tsinghua University under Grant No.2021GQG0001, in part by the Special Program of National Natural Science Foundation of China under Grant No.T2341003, in part by the Major Program of the National Social Science Fund of China under Grant No.21ZD19, in part by the Advanced Discipline Construction Project of Beijing Universities, and in part by the Nation Culture and Tourism Technological Innovation Engineering Project (Research and Application of 3D Music).
\end{acks}

\bibliographystyle{ACM-Reference-Format}
\bibliography{myReference}

\clearpage

\end{document}